\documentclass[sigconf]{acmart}
\AtBeginDocument{%
  }

\copyrightyear{2025}
\acmYear{2025}
\setcopyright{acmlicensed}\acmConference[SIGIR '25]{Proceedings of the 48th International ACM SIGIR Conference on Research and Development in Information Retrieval}{July 13--18, 2025}{Padua, Italy}
\acmBooktitle{Proceedings of the 48th International ACM SIGIR Conference on Research and Development in Information Retrieval (SIGIR '25), July 13--18, 2025, Padua, Italy}
\acmDOI{10.1145/3726302.3729922}
\acmISBN{979-8-4007-1592-1/2025/07}

\usepackage{multirow} 

\begin{document}
\begin{sloppypar}
\title{Clarifying Ambiguities: on the Role of Ambiguity Types in Prompting Methods for Clarification Generation}

\author{Anfu Tang}
\orcid{0000-0002-2336-1330}
\authornote{Corresponding author.}
\affiliation{%
  \institution{Sorbonne Université, CNRS, ISIR}
  \city{F-75005 Paris}
  \country{France}
}
\email{tang@isir.upmc.fr}

\author{Laure Soulier}
\orcid{0000-0001-9827-7400}
\affiliation{%
  \institution{Sorbonne Université, CNRS, ISIR}
  \city{F-75005 Paris}
  \country{France}}
\email{laure.soulier@isir.upmc.fr}

\author{Vincent Guigue}
\orcid{0000-0002-1450-5566}
\affiliation{%
  \institution{AgroParisTech, UMR MIA-PS}
  \city{Palaiseau}
  \country{France}
}
\email{vincent.guigue@agroparistech.fr}

\renewcommand{\shortauthors}{Anfu Tang, Laure Soulier, \& Vincent Guigue}


\begin{abstract}
  In information retrieval (IR), providing appropriate clarifications to better understand users' information needs is crucial for building a proactive search-oriented dialogue system. Due to the strong in-context learning ability of large language models (LLMs), recent studies investigate prompting methods to generate clarifications using few-shot or Chain of Thought (CoT) prompts. However, vanilla CoT prompting does not distinguish the characteristics of different information needs, making it difficult to understand how LLMs resolve ambiguities in user queries. In this work, we focus on the concept of ambiguity for clarification, seeking to model and integrate ambiguities in the clarification process. To this end, we comprehensively study the impact of prompting schemes based on reasoning and ambiguity for clarification. The idea is to enhance the reasoning abilities of LLMs by limiting CoT to predict first ambiguity types that can be interpreted as instructions to clarify, then correspondingly generate clarifications. We name this new prompting scheme \textsc{Ambiguity Type-Chain of Thought} (\textsc{AT-CoT}). Experiments are conducted on various datasets containing humanannotated clarifying questions to compare \textsc{AT-CoT} with multiple baselines. We also perform user simulations to implicitly measure the quality of generated clarifications under various IR scenarios.
\end{abstract}




\maketitle

\section{Introduction}
Ambiguity in information retrieval (IR) is a common factor that could undermine the quality of the retrieved documents. Indeed, real-world users often provide ambiguous queries to initialize a search without further elaboration~\citep{BelkinSigir03}. The reasons for this ambiguity can vary~\citep{BelkinSigir03,MorrisChi08}, such as avoiding the effort of typing lengthy queries, uncertainty about information needs, the tip-of-the-tongue phenomenon~\citep{arguello2021tip}, etc. The ambiguity of natural language itself could also account for this ambiguity in user queries, such as synonyms or polysemies~\citep{krovetz1997homonymy}. Regardless of the different causes, ambiguity is essentially a form of uncertainty, i.e. we cannot discern users' real intents by a single query. To better understand the ambiguities underlying user queries, previous studies have investigated ambiguity types (ATs) and proposed different taxonomies~\citep{min-etal-2020-ambigqa,guo2021abgcoqa,clarke2009overview} to classify ambiguities. 

\begin{figure*}[ht]
  \includegraphics[width=0.9\textwidth]{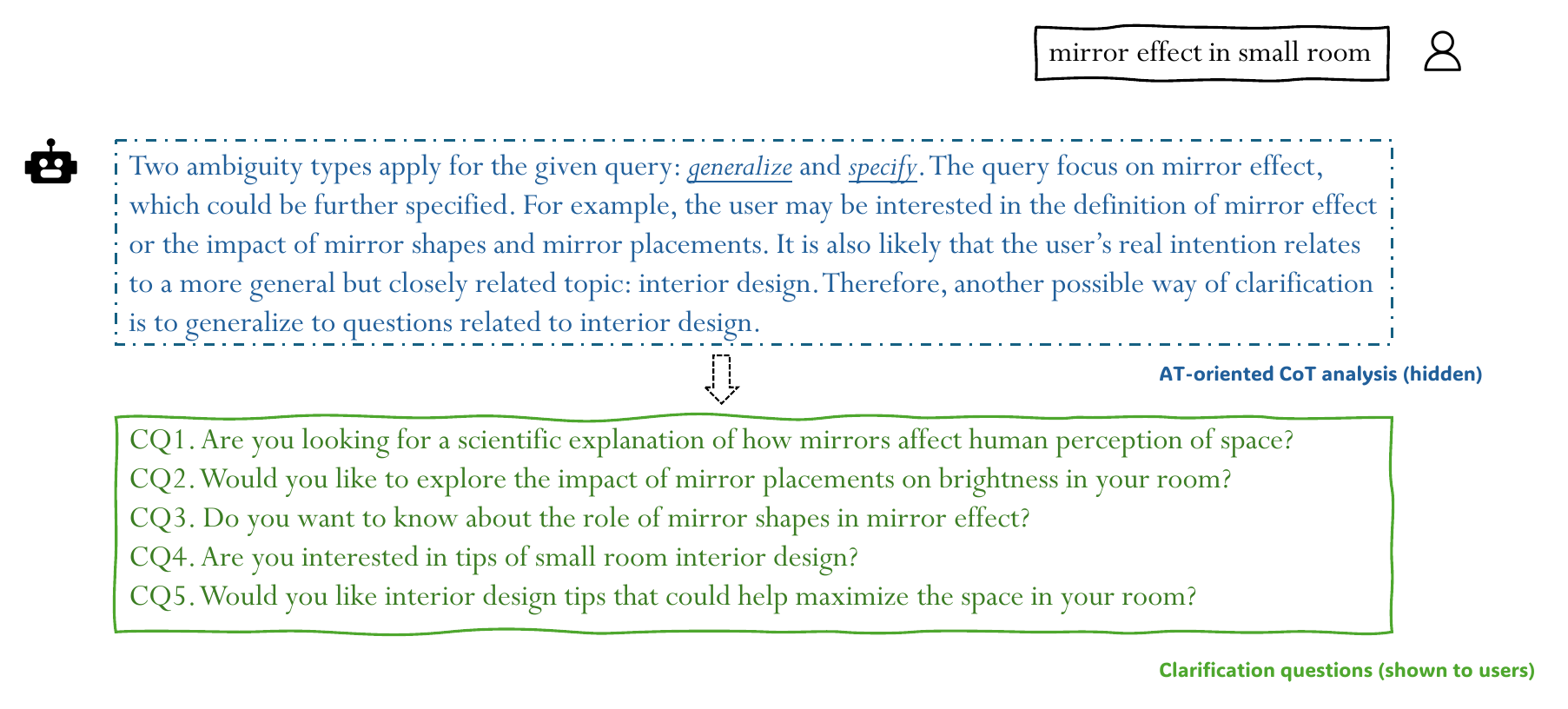}
  \vspace{-0.2cm}
  \caption{Illustration of \textsc{AT-CoT}: Unlike vanilla CoT, LLM-generated reasoning is limited to predict possible ambiguity types.}
  \Description{Illustration of AT-CoT: Unlike vanilla CoT, LLM-generated reasoning is limited to predicting possible ambiguity types.}
  \label{fig:teaser_conversation}
\end{figure*}

To navigate users through the ambiguity, we need a method that facilitates users in expressing their needs without compromising their user experience. Previous studies seek to achieve this goal by building proactive search-oriented dialogue systems~\citep{RadlinskiChiir17}, which can take the initiative to provide information or suggestions to help improve the quality of search results, including providing clarifications. Instead of passively receiving a list of documents, users can actively participate in the search by communicating with the proactive dialogue system through conversations. Early studies on clarifications focus on reformulated queries~\citep{carpineto2012survey,kuzi2016query}, which seek to provide useful suggestions that may meet the user's need, without explicitly exploiting the user's intent. Recent studies focus more on asking clarifying questions ~\citep{AliannejadiSigir19,rahmani-etal-2023-survey}, which consists of providing a clarifying question and allowing the user to respond freely. Clarification generation methods have evolved with the development of large language model (LLM), from supervised methods that rely on human-annotated data~\citep{guo2021abgcoqa,Amplayo2023QueryRP}, to LLM prompting methods~\citep{wangACM23,zhang-etal-2024-clamber}, among which Chain of Thought (CoT)  prompting ~\citep{wei2022chain} is found to generate better clarifying questions ~\citep{deng-etal-2023-prompting,zhang-etal-2024-clamber} compared to prompts with no generated reasoning. However, previous work mostly uses CoT prompts to freely generate reasoning, without explicitly asking LLMs to distinguish different information needs. We argue that understanding and integrating ambiguities into reasoning is important for the clarification process, since humans may first categorize the scenario of ambiguities, and then decide how to clarify the query properly. To simulate how humans handle ambiguous queries, we seek to first analyze the concept of ambiguity from the perspective of ambiguity types, and then integrate them into reasoning for clarification generation. To achieve this, we combine ambiguity types with CoT prompting to build \textsc{Ambiguity Type-Chain of Thought} (\textsc{AT-CoT}), which prompts LLMs to predict Ambiguity Types (ATs) that underlie a given query before generating clarifications correspondingly. To make \textsc{AT-CoT} properly work, we distill an action-based AT taxonomy from existing studies. 
Each AT in our taxonomy serves not only the purpose of helping LLMs understand ambiguity causes, but can also be interpreted as an instruction for LLMs to generate clarifications. Figure~\ref{fig:teaser_conversation} illustrates \textsc{AT-CoT}. Provided a query \textit{mirror effect in small room}, our method first predicts two ATs, then performs the corresponding actions to generate clarifying questions (CQs): CQ4 and CQ5 \textit{generalize} the query; CQ1, CQ2, and CQ3 \textit{specify} the query.   

To validate the effectiveness of our method, experiments are carried out on both intrinsic and extrinsic tasks (resp. clarification generation and IR) on numerous datasets including Qulac \citep{AliannejadiSigir19}, ClariQ \citep{aliannejadi2021building}, and TREC IR collections \citep{clarke2009overview,clarke2010Overview,clarke2011overview,clarke2012Overview}. For the IR task, following previous work \citep{AliannejadiSigir19,zouAcmTransInfSyst23,erbacher2024augmenting}, we perform user simulation to generate multi-turn conversations and then transform the generated conversations to reformulated queries. We compare different clarification interaction scenarios such as proposing query reformulations for users to select (\textit{select}) and asking a single clarifying question for users to respond (\textit{respond}). To summarize, our main contribution is as follows: 

$\bullet$ We analyze ambiguities from the perspective of ambiguity types, comprehensively investigate the impact of integrating ambiguities and reasoning in LLM prompting methods for
clarification.

$\bullet$ We validate the effectiveness of our method through experiments on clarification generation and IR tasks.

\section{Related Work}
\subsection{Ambiguity in User Queries}\label{sec:previous_at_studies}
\begin{figure}[t]
    \centering
    \includegraphics[width=1.1\linewidth]{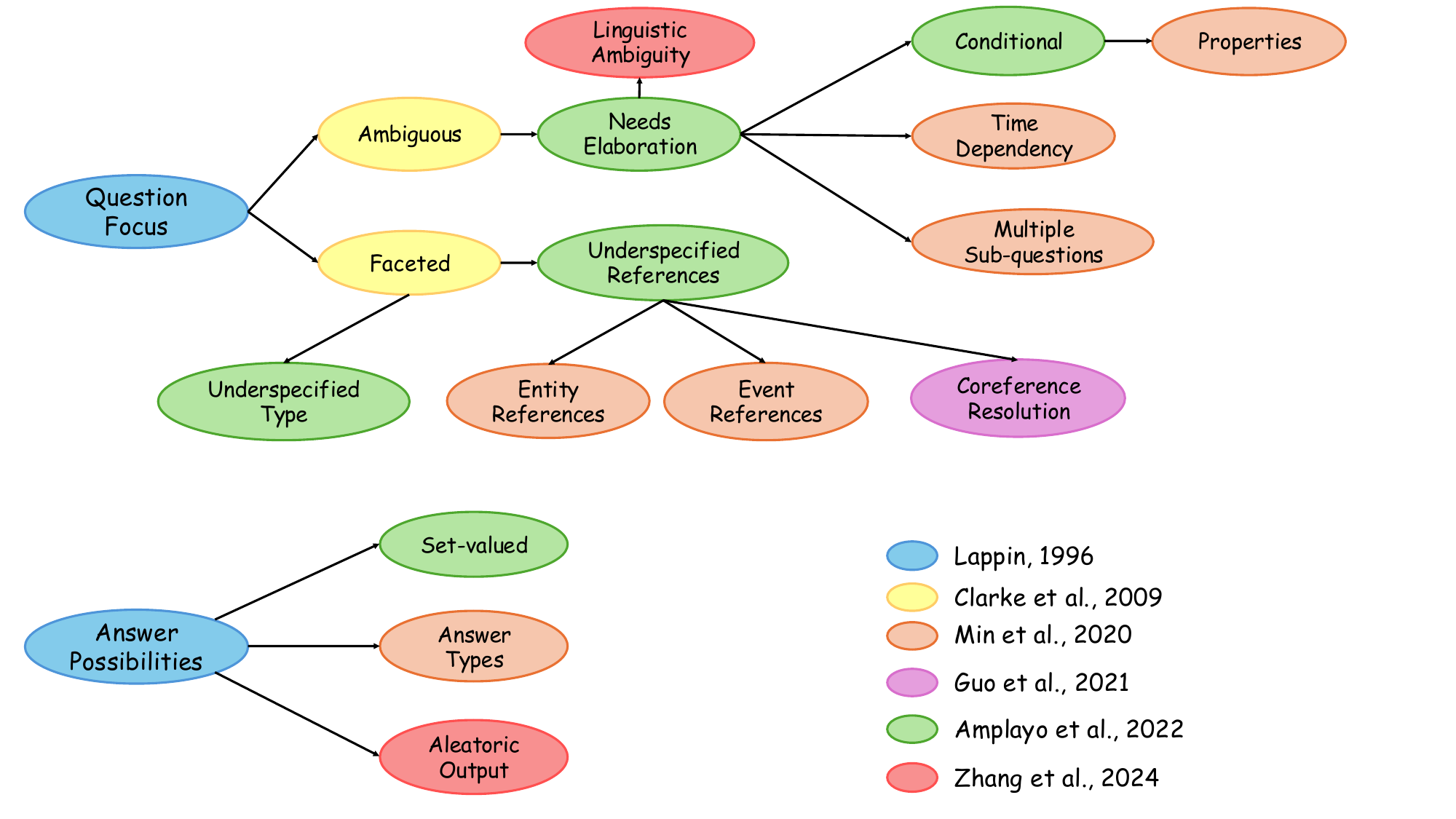}
    \caption{Ambiguity type taxonomies for analytical purposes in previous work.}
    \label{fig:comparison_between_ambiguity_type_taxonomy}
\end{figure}

While there is a lack of a widely accepted taxonomy of ambiguities, ambiguous queries have been long studied in the IR community \citep{clarke2009overview,zamani2020generating,zhang-etal-2024-clamber}. Previous work on ambiguity types (ATs) can be categorized into three types. The first group of studies formulates an AT taxonomy by analyzing queries in specific datasets \citep{clarke2009overview,guo2021abgcoqa,Amplayo2023QueryRP,min-etal-2020-ambigqa}. For instance, \citet{guo2021abgcoqa} proposed a taxonomy based on ambiguous questions in Abg-CoQA \citep{guo2021abgcoqa} with four ambiguity types: \textit{Coreference resolution} (unclear reference of pronouns), \textit{Time-dependency} (the interpretation of question depends on time), \textit{Answer types} (multiple answer possibilities) and \textit{Event references} (an entity in the question corresponds to multiple events). However, taxonomies in these studies are proposed more for analytical purposes and contain very specific ATs (e.g. \textit{EntityReferences} \citep{min-etal-2020-ambigqa} and \textit{CoreferenceResolution} \citep{guo2021abgcoqa} both correspond to a specific type of semantic ambiguity). Unlike these studies, we seek in this paper to formulate an AT taxonomy containing mutually exclusive ATs that can help LLMs generate better clarifications, rather than analyzing in detail why a query is ambiguous. Another group of studies focuses on the relations between queries by mining query logs \citep{zamani2020generating,jansen2009patterns,boldi2011query,lau1999patterns}, mostly based on sampled query reformulations from query logs. Although these studies may not be directly related to clarification generation, their findings provide useful insights into clarification patterns. For instance, two common query reformulation patterns observed \citep{jansen2009patterns,boldi2011query} are \textit{Generalization} and \textit{Specialization}. While the latter is widely considered in studies related to clarification, the need for generalization is less investigated. We argue that generalization can also help specify the information needs of users in certain scenarios. It corresponds to an important dimension of ambiguity, reflecting the possibility that user queries may fail to accurately convey user intent. The last group consists of studies in the post-LLM era. We notice a recent work \citep{zhang-etal-2024-clamber} that proposed a well-organized taxonomy with a special focus on ambiguities specific to LLMs, such as misaligned interpretations of queries between LLMs and humans. Our work differs from theirs: their taxonomy is used more as a tool to evaluate LLM performances in handling ambiguous queries, while our work focuses on integrating ambiguity types into reasoning for clarification generation. In a nutshell, previous work mostly exploit ATs for analysis, without searching to enhance the reasoning ability of LLM prompting by integrating ATs. To help better understand previous work, we organize ATs in existing taxonomies and present them in Figure~\ref{fig:comparison_between_ambiguity_type_taxonomy}.

\subsection{Clarification in Information Retrieval} 
Clarification serves the purpose of eliciting the user's information need \citep{zamani2020generating} by exchanging with the user and exploiting the user's feedback. The clarification form could be diverse, either by proposing reformulated queries to diversify the retrieval results or by asking clarifying questions to induce users to clarify. Early attempts of clarification generation focus on automatic query expansion  \citep{carpineto2012survey}, whereby users' original queries are rewritten or augmented. For example, \citet{ChiritaSigir07} proposed to expand user queries with terms collected from user data to handle ambiguities of short keyword queries in web searches. Other studies \citep{cucerzan2007query,CaoKDD08,MeiCIKM08} investigate query suggestions by exploring different user-specific data sources such as landing pages, clicks, or hitting time. Both query expansion and query suggestions can be regarded as a form of proposing reformulated queries to users and collecting users' feedback. Recent studies concentrate more on asking clarifying questions \citep{rao-daume-iii-2018-learning,rahmani-etal-2023-survey,AliannejadiSigir19,xu-etal-2019-asking,lee-etal-2023-asking}. The common approach consists of using the conversation history as input to a generative language model to generate CQs.

Before the era of LLM, clarification generation methods mostly consist of training sequence-to-sequence neural models (e.g. seq2seq \citep{sutskeverNIPS14}) using labeled data. For example, \citet{guo2021abgcoqa} fine-tune BART \citep{lewis-etal-2020-bart} to generate clarifying questions with ambiguous questions provided as input; \citet{xu-etal-2019-asking} investigated knowledge-based clarifying question generation and concatenated entity texts and the current question as input to a Seq2seq \citep{Bahdanau2014NeuralMT} model. Recent studies in the post-LLM era have increasingly focused on LLM prompting methods, such as using few-shot prompting~\citep{lee-etal-2023-asking,zhang-etal-2024-clamber}, Chain of Thought (CoT) prompting~\citep{deng-etal-2023-prompting,zhang-etal-2024-clamber}. Our work extends existing studies on LLM prompting methods for clarification generation by integrating ambiguity types into CoT reasoning.

\subsection{Conversation Simulation in Search}
In IR, user simulation consists of creating artificial conversations based on hypotheses about user behaviors, often used to automatically test the performance of dialogue systems without performing real user tests \citep{eckert1997user,dupret2008user,chuklin2013modeling}. The common approach is to instantiate a user agent to communicate with the dialogue system according to certain strategies~\citep{erbacher2022interactive,camara2022ECIR,maxwell2015searching}. Hypotheses about user behavior are made to control how user agents respond, depending on the purpose of the simulation. For example, to test the quality of reformulation in IR systems, \citet{erbacher2022interactive} assumed that the user agent is greedy and fully cooperative, thus always selecting the reformulations most similar to the user intent. In another work \citep{erbacher2024augmenting}, user agents are allowed to only respond 'yes' or 'no' to augment IR datasets with multi-turn conversations. Some studies involve simulation of more complex user behaviors, in which user agents are initialized with different variables, each corresponding to a specific type of user. Recent studies have increasingly focused on LLM-based conversation simulation. For example, \citet{owoichoSIGIR23} built a user simulator for mixed-initiative multi-turn conversation systems by prompting LLMs. Following previous work on LLM-based conversation simulation in IR, in our work, we instantiate our user agent using LLMs and simulate user responses by few-shot LLM prompting.

\begin{table*}[ht]
    \centering
    \caption{Proposed action-based ambiguity type.}
    \begin{tabular}{p{3cm}|p{7cm}|p{5cm}}
        \hline
        \textbf{Ambiguity Type} & \textbf{Definition} & \textbf{Related ATs from previous work} \\ 
        \hline
        \textit{Semantic} & The query is semantically ambiguous for several common reasons: it may include homonyms; a word in the query may refer to a specific entity while also functioning as a common word; or an entity mentioned in the query could refer to multiple distinct entities. & \textit{Question Focus}~\citep{Lappin1996-LAPTHO} \newline \textit{Linguistic Ambiguity}~\citep{zhang-etal-2024-clamber} \\
        \hline
        \textit{Generalize} & The query focuses on specific information; however, a broader, closely related query might better capture the user's true information needs. & \textit{Generalization}~\citep{jansen2009patterns,boldi2011query} \\
        \hline
        \textit{Specify} & The query has a clear focus but may encompass too broad a research scope. It is possible to further narrow down this scope by providing more specific information related to the query. & \textit{Faceted}~\citep{clarke2009overview} \newline \textit{Time Dependency}~\citep{min-etal-2020-ambigqa} \newline \textit{Underspecified References}~\citep{Amplayo2023QueryRP} \\
        \hline
    \end{tabular}
    \label{tab:ambiguity_types}
\end{table*}

\section{Methodology}
We present here the outline of our methodology. We focus on the following research questions:
\begin{enumerate}
    \item[\textbf{RQ1.}] What is an appropriate taxonomy of ambiguities for generating clarifying questions that is compatible with LLM prompting methods?
    \item[\textbf{RQ2.}] How to integrate ambiguity and reasoning in LLM prompting methods for clarification?
\end{enumerate}

\subsection{Ambiguity Type Taxonomy}\label{sec:proposed_at_taxonomy}
To respond to RQ1, we first seek to exploit ambiguity types to concretize the concept of ambiguity. The goal is to establish a taxonomy that can be used to enhance LLM reasoning ability in terms of handling ambiguous queries. Previous work \citep{min-etal-2020-ambigqa,guo2021abgcoqa,Amplayo2023QueryRP} proposed various ambiguity-type taxonomies for the analytical purpose. However, from the perspective of helping LLMs understand ambiguities and better instructing LLMs to generate clarifications, we find existing taxonomies redundant and unsuitable for LLM prompting methods. Firstly, as evidenced by ~\citet{zhang-etal-2024-clamber}, existing taxonomies were mostly proposed before the era of LLMs, some ATs lack clear definitions and ATs are not mutually exclusive. Secondly, ATs in existing taxonomies can be reduced to two actions that LLMs can take: \textit{Determine the Query Interpretation} or \textit{Further Specify the User Query}. Following ~\citet{deng-etal-2023-prompting} who proposed proactive prompting, i.e. making LLMs decide actions to take instead of simply responding to instructions, we propose an LLM action-based taxonomy that encompasses three dimensions, each corresponding to a clarification pattern discovered in previous work~\citep{clarke2009overview,zamani2020generating,jansen2009patterns}: 

\begin{itemize}
    \item \textit{Semantic}: accounts for ambiguity in query interpretations.  
    \item \textit{Generalize}: addresses ambiguity in information needs when users seek relevant yet more general information. It occurs when user queries do not precisely describe real user intents.  
    \item \textit{Specify}: addresses ambiguity in information needs when users seek more specific information. It occurs when user queries lack details and may correspond to a too large search scope. Most ATs in existing taxonomies can be categorized under this category (e.g. \textit{NeedsElaboration}, \textit{UnderspecifiedReferences}~\citep{Amplayo2023QueryRP}). 
\end{itemize}

Table~\ref{tab:ambiguity_types} presents detailed explanations of our AT taxonomy. Compared to previous taxonomies, the strength of our taxonomy lies in its dual function: each AT in our taxonomy not only helps LLMs understand underlying ambiguities, but can also be easily interpreted as an action for LLMs to take.

\begin{table*}[h]
    \centering
    \caption{Prompts of four prompting schemes: standard, AT-standard, CoT and AT-CoT. <AT definitions> is a placeholder for AT definitions in Table~\ref{tab:ambiguity_types}.}\label{tab:prompt_system_instruction}
    \resizebox{0.87\textwidth}{!}{\begin{tabular}{@{}lp{14cm}@{}}
        \toprule
        \textbf{Prompt Type} & \textbf{System Instruction} \\ 
        \midrule
        standard & \texttt{Given a query in an information-seeking system, generate a clarifying question that you think is most appropriate to gain a better understanding of the user's intent.
        \newline <query>} \\
        \midrule
        AT-standard & \texttt{Given a query in an information-seeking system, generate a clarifying question that you think is most appropriate to gain a better understanding of the user's intent. The ambiguity of a query can be multifaceted, and there are multiple possible ambiguity types:
        \newline <AT definitions>
        \newline Consider the above ambiguity types when generating. 
        \newline <query>} \\
        \midrule
        CoT & \texttt{Given a query in an information-seeking system, generate a clarifying question that you think is most appropriate to gain a better understanding of the user's intent.
        \newline Before generating the clarifying question, provide a textual explanation of your reasoning about why the original query is ambiguous and how you plan to clarify it. 
        \newline <query>} \\
        \midrule
        AT-CoT & \texttt{Given a query in an information-seeking system, generate a clarifying question that you think is most appropriate to gain a better understanding of the user's intent. The ambiguity of a query can be multifaceted, and there are multiple possible ambiguity types: 
        \newline <AT definitions>
        \newline Before generating the clarifying question, provide a textual explanation of your reasoning about which types of ambiguity apply to the given query. Based on these ambiguity types, describe how you plan to clarify the original query.  
        \newline <query>} \\
        \bottomrule
    \end{tabular}}
\end{table*}

\subsection{Prompting Formulation for Clarification Generation}\label{sec:prompting_formulation}
This section aims to respond to RQ2, i.e. how to integrate ambiguities, abstracted by the ambiguity type taxonomy in Section~\ref{sec:proposed_at_taxonomy}, into reasoning for LLM prompting methods. We aim to achieve this by constraining the reasoning of CoT prompting. Intuitively, we seek to require LLMs to predict ATs in our taxonomy to integrate ambiguities into reasoning, which endows LLMs the capability to reason in a way that we expect and take explainable actions to clarify ambiguous queries. We hypothesize that ambiguity-oriented reasoning is better than freely generated LLM reasoning. Therefore, we propose \textsc{Ambiguity Type-Chain of Thought} (\textsc{AT-CoT}) that extends CoT prompting. To effectively access the impact of integrating ambiguities into LLM reasoning, we use another two prompting schemes as baselines: standard prompting, which simply requires LLMs to generate clarifications without any intermediate steps; AT-standard prompting, for which we add definitions of ATs in our taxonomy into prompt instructions. We use AT-standard to validate the impact of simply informing LLMs of possible ATs without asking LLMs to generate reasoning. Table~\ref{tab:prompt_system_instruction} shows detailed system instructions for different prompting schemes. Mathematically, each prompting method can be formulated as follows: 

$\bullet$ \textbf{standard}~\citep{deng-etal-2023-prompting}: Standard prompting relies only on inherent knowledge of LLMs to generate clarifications without intermediate steps. The objective of standard prompting is to maximize:
\begin{equation}
    p(c|\mathcal{D},\mathcal{C},q)
\end{equation}
where $c$ denotes the generated clarification, $q$ denotes an ambiguous query, $\mathcal{C}$ denotes the conversation history, and $\mathcal{D}$ denotes the task description.

$\bullet$ \textbf{AT-standard}: The only difference between AT-standard and standard prompting is that AT definitions are included in the prompt:
\begin{equation}
    p(c|\mathcal{D},\mathcal{A},\mathcal{C},q)
\end{equation}
where $\mathcal{A}$ refers to the AT definitions from Table~\ref{tab:ambiguity_types}.

$\bullet$ \textbf{CoT (Chain of Thought)}~\citep{wei2022chain}: CoT prompting requires LLMs to generate texts of reasoning before making clarifications (reasoning without constraints):
\begin{equation}
    p(a,c|\mathcal{D},\mathcal{C},q)
\end{equation}
where $a$ refers to the generated textual reasoning.
    
$\bullet$ \textbf{AT-CoT}: AT-CoT requires LLMs to first predict ATs from our taxonomy, then generate clarifications correspondingly. The objective of AT-CoT prompting is to maximize:
\begin{equation}
    p(a,c|\mathcal{D},\mathcal{A},\mathcal{C},q)
\end{equation}

\section{Experimental Setup}\label{sec:experiment_setup}
To evaluate the effectiveness of \textsc{AT-CoT}, we conduct experiments on three types of tasks: (1) Clarification generation (CG), for which we use datasets containing human-annotated clarifying questions (CQs) and we evaluate by computing the semantic similarity between generated CQs and human-annotated ones. (2) Information retrieval (IR), for which we simulate multi-turn conversations and transform conversations into reformulated queries to retrieve documents. (3) CG+IR, for which we align CG performance and IR performance to investigate the correlation between the performance of CG and IR, i.e., if better clarifications could improve IR performance. 

\subsection{Datasets}
We present datasets that are used in our experiments in this section. Table~\ref{tab:dataset_overview} summarizes the statistics of different datasets. 

\subsubsection{CG Datasets}~\\
$\bullet$ Qulac \citep{AliannejadiSigir19}: Qulac uses queries from TREC web track 2009-2012. Annotators are asked to first figure out facets related to given queries by scanning snippets of web searching results using a search engine, then generate CQs to address the facets.\\  
$\bullet$ ClariQ \citep{aliannejadi2021building}: Similarly to Qulac, ClariQ is crowdsourced by annotating CQs for provided queries. Ambiguity level labels ranging from 1-4 are provided in ClariQ, with 4 representing extreme ambiguous queries. \\
$\bullet$ RaoCQ \citep{rao-daume-iii-2018-learning}: A domain-specific dataset containing clarifying question annotations. Annotators are asked to identify relevant CQs given (question, follow-up questions) pairs, where each question refers to an original question from a post on StackExchange, and follow-up questions are sampled from follow-up questions in comments of the same post. Similarly to \citep{rao-daume-iii-2018-learning}, in this work, we evaluate our methods on the subset with human annotations.\\

\footnotetext{\url{{https://lucene.apache.org/}}}

\subsubsection{IR Datasets}~\\
$\bullet$  TREC Web track 2009-2012~\citep{clarke2009overview,clarke2010Overview,clarke2011overview,clarke2012Overview}: An IR dataset that focus on web search queries. We use ClueWeb09\footnote{\url{https://lemurproject.org/clueweb09.php/}} Category B as the document collection which contains 50 million English web pages. Since facet-specific document relevance judgments are provided in TREC web track diversity tasks, we use facets as user intents in user simulation.  ~\\
$\bullet$  TREC Web track 2013-2014~\citep{collins2013Overview,collins2014Overview}: As TREC Web track 2009-2012, TREC Web track 2013-2014 contains multifaceted web search queries, while including more focused topics to present more challenging queries. ClueWeb12\footnote{\url{https://lemurproject.org/clueweb12/}} is used as the document collection. ~\\
$\bullet$ TREC DL Hard~\citep{mackie2021deep}: A benchmark containing with queries from TREC DL 2019 \& 2020~\citep{Craswell2019TrecDl,Craswell2020TrecDl}, which we believe may require multi-turn clarification to resolve implied ambiguities. The queries in TREC DL Hard are sampled from MS Marco~\citep{Campos2016MSMA}. We use the MS Marco passage corpus as the document collection. 

\subsubsection{CG+IR Dataset}\label{sec:cg_ir_datasets}
Since Qulac is based on queries from TREC Web Track 2009-2012, we align Qulac queries to document relevance judgments provided by TREC Web Track 2009-2012. We refer to this dataset by Qulac-TREC Web Track 2009-2012, which contains human-annotated CQs from Qulac and document relevance judgments from TREC Web Track 2009-2012.

\begin{table}[t]
    \centering
    \caption{Statistics of datasets used in our experiments for three tasks: CG, IR, CG+IR.}
    \resizebox{\linewidth}{!}{\begin{tabular}{lccc}
        \hline
        \textbf{Dataset} & \textbf{\# queries} & \textbf{\# CQs} & \textbf{\# intents}\\ 
        \hline
        \multicolumn{4}{c}{ \textit{Task 1: CG}} \\
        \hline
        \textit{Qulac} & 198 & 2575 & - \\
        \textit{ClariQ} & 298 & 3991 & - \\
        \textit{RaoCQ} & 500 & 2248 & - \\
        \hline
        \multicolumn{4}{c}{ \textit{Task 2: IR}} \\
        \hline
        \textit{TREC Web Track 2009-2012} & 198 & - & 717 \\
        \textit{TREC Web Track 2013-2014} & 100 & - & 315 \\
        \textit{TREC DL Hard} & 50 & - & 350 \\
        \hline
        \multicolumn{4}{c}{ \textit{Task 3: CG+IR}} \\
        \hline
        \textit{Qulac-TREC Web Track 2009-2012} & 198 & 2575 & 717 \\
        \hline
    \end{tabular}}
    
    \label{tab:dataset_overview}
\end{table}

\subsection{Evaluation Protocol}\label{sec:evlauation_protocol}
\paragraph{Clarification Generation.} For each query, we generate multiple CQs to fairly evaluate the performance of different prompting methods on the clarification generation (CG) task. Several factors drive this decision: 1) In CG datasets, each query corresponds to numerous human-annotated CQs, covering different clarification possibilities. However, human-annotated CQs cannot contain all possible CQs. Since automatic metrics such as BERTScore~\citep{bert-score} capture the semantic similarity between generated CQs and reference CQs, it is likely that a high-quality generated CQ gets a low BERTScore. To mitigate this issue, we seek to generate multiple diverse CQs, therefore reducing the probability that none of the generated CQs is semantically similar to any of the human-annotated ones. 2) For \textsc{AT-CoT}, since multiple ambiguity types may exist for a query, it is natural to generate multiple CQs to account for different ATs. For other prompting methods that do not predict ATs, generating multiple CQs is also helpful for fair comparison. As evidenced in \citet{Wang2022SelfConsistencyIC}, the voting strategy can increase the performance of LLM prompting methods. Generating multiple CQs and comparing the best-performing CQ can be regarded as a type of voting, by which we reduce the variance of prompting performances on CG.

\begin{figure*}[h]
    \centering
    \includegraphics[width=0.9\textwidth]{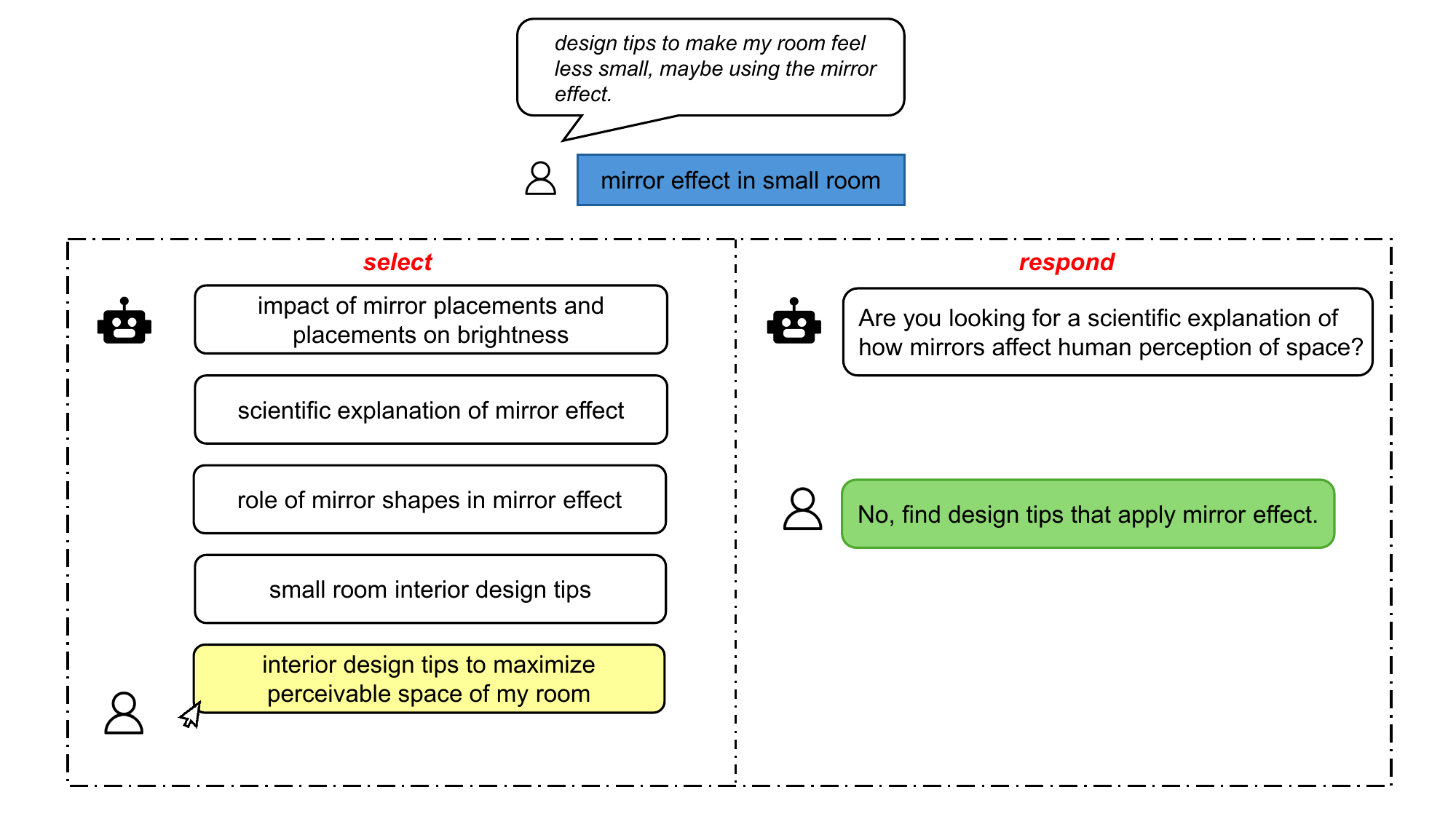}
    \caption{Illustration of the two clarification interaction scenarios in our user simulation: \textit{select} and \textit{respond}.}
    \label{fig:clarification_form_illustration}
\end{figure*}

\paragraph{User Simulation for IR} We also evaluate the impact of our methodology on IR via user simulation. Two clarification interaction scenarios are tested (Figure~\ref{fig:clarification_form_illustration}):

$\bullet$ \textit{select}: In each turn, 5 RQs are generated. We adopted a moderate temperature of 0.6 to balance the diversity and consistency of the generated RQs, ensuring that they are varied while avoiding excessive creativity. The user agent selects the RQ that best corresponds to their intent, and the conversation continues based on the selected RQ.

$\bullet$ \textit{respond}: In each turn, a CQ is generated. The user agent responds to it based on the provided user intent.

We choose the two interaction scenarios for the following reasons: 1) \textit{select} is widely studied in previous studies~\citep{cucerzan2007query,CaoKDD08}, and applied in certain real-world scenarios such as search engine suggestions through query reformulation. 2) \textit{respond} corresponds to the more naturalistic settings for conversational search modeling interactions in natural language \citep{AliannejadiSigir19,lee-etal-2023-asking,SekulicACM21}.  A baseline w/o clarification is also considered, which uses original queries without clarification for document retrieval.

To simulate multi-turn conversations, three prompts are chained: \textit{generation}, \textit{response}, \textit{reformulation}. The \textit{generation} prompt has four variants, each representing a prompting methods as described in Section~\ref{sec:prompting_formulation}. We use this prompt to generate RQs (scenario \textit{select}) or CQs (scenario \textit{respond}). The \textit{response} prompt generates simulated user responses and has two variants, each corresponding to an interaction scenario. The clarifications generated by \textit{generation} and paragraphs describing user intents are used as input for the \textit{response} prompt. Since facet-specific document relevance judgments are provided in TREC Web track 2009-2014, we directly use facets as user intents for \textit{response} and use gold document relevance labels. For TREC DL Hard, we use each relevant document as user intent for \textit{response}, then use the same document as the gold label in IR evaluation. For each simulated conversation, the \textit{reformulation} prompt uses the simulated conversation as input and summarizes the conversation into a reformulated query. The objective of using \textit{reformulation} is to facilitate the evaluation of IR tasks, since most existing IR models retrieve documents by queries rather than conversations. Examples of the \textit{response} and \textit{reformulation} prompts can be found in Table~\ref{tab:respond_reformulation_prompts}. As a result, we have eight types of simulated conversations, each corresponding to a unique (clarification generation prompt, interaction scenario) combination. To simplify the simulation, we assume that users are always cooperative and that no intent shifts occur during the conversation. Each conversation is initialized by a user query and simulated to three turns with no stopping rules. User agents are always provided with complete descriptions of user intents.

\begin{table*}
    \centering
    \caption{Prompts of \textit{respond} and \textit{reformulation}. There is no \textit{reformulation} prompt for the mode \textit{select}.}\label{tab:respond_reformulation_prompts}
    \scalebox{0.9}{
    \begin{tabular}{@{}lp{14cm}@{}}
        \toprule
        \textbf{Prompt type} & \textbf{System instruction} \\ 
        \midrule
        \textit{response} (scenario \textit{select}) & \texttt{Imagine that you are a user seeking information with the help of a conversational assistant. At each turn of the conversation, the assistant provides you several reformulated queries to better understand your intent. Given a conversation history and a paragraph describing the user intent, choose the reformulated query that most accurately reflects the provided user intent.}
        \newline \texttt{<chat history>}
        \newline \texttt{<user intent>}\\
        \midrule
        \textit{response} (scenario \textit{respond}) & \texttt{Imagine that you are a user seeking information with the help of a conversational assistant. At each turn of the conversation, the assistant asks a clarification question to better understand your intent. Given a conversation history and a paragraph describing the user intent, respond to the clarification question based on the provided user intent.}
        \newline \texttt{<chat history>}
        \newline \texttt{<user intent>}\\
        \midrule
        \textit{reformulation} (scenario \textit{respond}) & \texttt{Given a conversation history, summarize the conversation as a reformulated query. The conversation history includes the initial query and several clarification turns between the user and a virtual assistant.} 
        \newline \texttt{<chat history>}
        \\
        \bottomrule
    \end{tabular}}
\end{table*}

\subsection{Evaluation Metric}
Following \citep{zhang-etal-2024-clamber}, we use BERTScore \citep{bert-score} for CG tasks, since metrics based on N-gram matching like BLEU or ROUGE cannot measure clarification abilities \citep{guo2021abgcoqa}. As mentioned in Section~\ref{sec:evlauation_protocol}, we ask LLMs to generate multiple CQs. For each query, we compute a score for generated CQs as follows: suppose that for each query $q$, we generate a list of CQs $(gcq_1,...,gcq_M)$, with a list of annotated CQs $(acq_1,...,acq_N)$ as gold standards. We first compute a query-specific score matrix $S$: 
\begin{equation}
    S_{i,j}=BERTScore(gcq_i,acq_j)
\end{equation} 

where $i=1,...,M$, $j=1,...,N$. The score on $q$ is computed by: $score_q=max(S_{[:,:]})$, which is the maximum value of $S$. We take the BERTScore of the best-performing generated CQ to access the overall CG performance on $q$ for the following reason: a CG method is good if it is able to generate a CQ that is highly similar to one of the reference CQs.

For the IR task, we use different standard metrics following previous work~\citep{clarke2009overview,zhouACM24}: We use nDCG@10 (Normalized Discounted Cumulative Gain~\citep{Wang2013ATA}) for TREC web track 2009-2014. For TREC DL Hard, since we use each relevant document as user intent for simulation then verify whether the target document is ranked higher through clarification, we use MRR@10 (Mean Reciprocal Rank~\citep{radev-etal-2002-evaluating}) as the evaluation metric.

\subsection{Implementation Details}\label{sec:implementation_details}
\paragraph{Prompting Scheme} Following previous work \citep{deng-etal-2023-prompting}, we adopt few-shot settings for all prompting schemes. We have two reasons to do so: 1) results of preliminary experiments demonstrate that few-shot prompting always significantly outperform zero-shot, regardless of the prompting scheme; 2) zero-shot prompting is likely to generate over lengthy analysis, causing incomplete generation due to maximum output token limitation or slowing down inference. In this work, we assume without further notice that all prompting methods are under few-shot settings.

\paragraph{LLM} We use Llama-3-8B \citep{dubey2024llama} as our base model and load pre-trained weights from Huggingface. LLM hyperparameters are fixed with: $k=10$ for top-$k$ sampling; temperature $t=0.6$. Due to the extensive prompting inference involved in our experiments, we quantize Llama-3 to NF4 (4-bit NormalFloat) and conduct our experiments on a single 12G TITAN Xp GPU.

\paragraph{Parsing LLM Outputs} We ask LLMs to give JSON-style structured outputs through format instructions and few-shot examples containing reference formatted outputs. LLM outputs are parsed using the Pydantic parser from LangChain\footnote{\url{https://python.langchain.com/v0.1/docs/modules/model_io/output_parsers/types/pydantic/}}. In case of parsing errors, we ask LLMs to regenerate the outputs with a maximum number of retry attempts set to 10. In rare cases, we manually parse the LLM outputs to address persistent parsing errors. 

\paragraph{BERTScore} We use the third-ranked pre-trained model of BERTScore\footnote{{\url{https://huggingface.co/microsoft/deberta-large-mnli}}} based on their experimental results \footnote{\url{https://github.com/Tiiiger/bert_score}} on the WMT16 machine translation task \citep{bojar2016findings}.

\paragraph{IR Pipeline} Following \citep{nogueira2019passage,karpukhin-etal-2020-dense,nogueira-etal-2020-document}, we adopt a two-stage retriever-rerank pipeline for IR tasks. Top-$k$ relevant documents are first retrieved from a large-scale document collection using BM25 and then reranked using MonoT5. For the retriever, we use a no-tuning \textit{pyserini}\footnote{\url{https://github.com/castorini/pyserini}} Lucene implementation with $k$ fixed to 100. For the reranker, we use a pre-trained MonoT5 \citep{nogueira-etal-2020-document}.

\paragraph{IR Datasets} We use \textit{ir\_datasets} \citep{macavaneyIrDatasets}, a commonly used Python package in IR community to manage IR datasets. The Python implementation of \textit{pyserini} is used to build BM25 indexes of Clueweb09 and Clueweb12.

\section{Task 1: Clarification Generation (CG)}
This section aims to evaluate the impact of integrating ambiguities into LLM reasoning on the performance of the clarification generation (CG) task. Table~\ref{tab:cg_results} depicts the overall comparison between different prompting methods using various datasets and the BERTScore metric. Results show that AT-CoT consistently outperforms the three baselines with significant margins across all datasets. For instance, AT-CoT reaches a BERTScore of 82 vs. scores ranging from 78.8 to 80 for other baselines on ClariQ. This suggests that ambiguity-oriented reasoning (AT-COT) helps generate better clarifying questions. This improvement is consistent on both specific-domain datasets (RaoCQ) and open-domain datasets (Qulac, ClariQ), showing that our method generalizes to different types of queries. Besides, through the comparison between AT-standard and standard prompting, we find that only informing LLMs of existing ambiguity types is not helpful, and even degrades the CG performance in some cases (e.g. 77 vs 77.9 for resp. AT-standard vs. standard on Qulac). This demonstrates that integrating ambiguity types is only helpful when integrated into LLM reasoning. Our observation on CoT prompting is coherent with previous work~\citep{deng-etal-2023-prompting,zhang-etal-2024-clamber}: CoT is more effective than standard prompting in terms of generating clarifying questions. We explore even further by claiming that ambiguity-oriented reasoning is more helpful. 

\renewcommand{\arraystretch}{1.2}
\begin{table}[b]
\setlength{\tabcolsep}{10pt}
\caption{Overall evaluation on CG datasets. $^{*}$, $^{\dag}$, $^{\Delta}$ marks statistically significant improvements over standard, AT-standard, CoT respectively with $p < 0.01$ under a t-test.}\label{tab:cg_results}
\vspace{-0,2cm}
\centering
\begin{tabular}{lccccccccccccccc}
\toprule
Prompt & Qulac & ClariQ & RaoCQ \\
\hline
standard & 77.9 & 79.3 & 60.0  \\

AT-standard & 77.0 & 78.8 & 59.9 \\

CoT & 79.2$^*$ & 80.0 & 60.5 \\

AT-CoT & \textbf{80.6}$^{*\dag\Delta}$ & \textbf{82.0}$^{*\dag\Delta}$ & \textbf{62.4}$^{*\dag\Delta}$ \\
\hline
\end{tabular}
\end{table}

\paragraph{Stratification by Ambiguity-level}
We further evaluate the performance of CG tasks across different ambiguity levels. We use labels provided in ClariQ for this analysis, and present results in Table~\ref{tab:cg_results_per_ambiguity_level}. Generally, both CoT and AT-CoT outperform standard and AT-standard prompting on the first three ambiguity levels, demonstrating the usefulness of both freely generated LLM reasoning and ambiguity-oriented reasoning for CG when queries are not extremely ambiguous. However, in cases of extreme ambiguity (level-4), the performance of CoT falls below standard prompting (BERTScore of 78 vs. 78.5 and 79.7 for standard and AT-standard resp.), meanwhile, the improvement of AT-CoT is still consistent (BERTScore=82.4). This suggests that ambiguity types could be particularly useful for handling ambiguous queries.

\begin{table}[h]
\setlength{\tabcolsep}{8pt}
\caption{CG results on ClariQ stratified by ambiguity levels. $^{*}$, $^{\dag}$, $^{\Delta}$ marks statistically significant improvements over standard, AT-standard, CoT, respectively.}\label{tab:cg_results_per_ambiguity_level}
\vspace{-0,2cm}
\centering
\begin{tabular}{cllll}
\hline
& level-1 & level-2 & level-3 & level-4 \\
\hline
standard & 78.7  & 80.0 & 78.9 & 78.5 \\

AT-standard & 77.6 & 79.2  & 78.4 & 79.7  \\

CoT & 78.6  & 80.5 & 80.7$^{\dag}$ & 78.0  \\

AT-CoT & \textbf{80.9}$^{*\dag}$ & \textbf{82.0}$^{*\dag\Delta}$ & \textbf{82.1}$^{*\dag\Delta}$ & \textbf{82.4}$^{\dag\Delta}$ \\
\hline
\end{tabular}
\end{table}

\paragraph{Distribution of Ambiguity Types} To provide more insight into AT-CoT, we analyze the distribution of ambiguity types predicted by AT-CoT. We first investigate the frequency of predicted ATs (namely \textit{Semantic}, \textit{Generalize} and \textit{Specify}), and then focus on the impact of predicting different ATs on the performance of the CG task. Predicted ATs are extracted from the reasoning generated by AT-CoT. Table~\ref{tab:ambiguity_type_distribution} shows statistics about each group on all CG datasets: the frequency of queries identified as a specific AT and the performance difference in terms of BERTScore between AT-CoT and CoT (in parentheses). Our main conclusions are the following: 1) \textit{Semantic} and \textit{Specify} are the most frequent types for all CG datasets, with \textit{Specify} being slightly more common (i.e. at most 45.9\% vs. at most 53.7\% for \textit{Specify}). This observation aligns with the fact that most ATs in existing taxonomies can be categorized as \textit{Semantic} or \textit{Specify}. However, though less frequent, the importance of \textit{Generalize} cannot be overlooked. 2) The \textit{Generalize} type is more marginal but the fact that 12\% of queries in RaoCQ are predicted to be generalizable justifies our decision to include \textit{Generalize} in our taxonomy. 3) The observation that queries in RaoCQ more often require generalization suggests that the AT predictions of AT-CoT effectively capture the clarification needs of queries and are less likely to be random. Since RaoCQ queries are extracted from user posts on StackExchange, they are generally longer compared to queries in Qulac and ClariQ. It is therefore very likely that a query in RaoCQ does not precisely describe user intents and requires generalization. Differently, queries in Qulac are often short, used for web search, making them less possibly to require generalization. This gap between the frequency of \textit{Generalize} being predicted and improvements caused by predicting \textit{Generalize} reflects that AT-CoT adapts well to datasets with different characteristics.

\begin{table}[t]
\caption{Distribution of ATs predicted by AT-CoT. In parentheses, we show corresponding CG performance differences between AT-CoT and CoT regarding the BERTScore.}\label{tab:ambiguity_type_distribution}
\centering
\begin{tabular}{p{1.5cm}ccc}
\hline
& Qulac & ClariQ & RaoCQ \\
\hline
\textit{Semantic}  & 44.6 ($\uparrow$ 1.3) & 45.9 ($\uparrow$ 1.8) & 42.4 ($\uparrow$ 2.0)  \\
\textit{Generalize} & 1.7 ($\downarrow$ 0.6) & 1.9 ($\uparrow$ 1.4) & 12.3 ($\uparrow$ 2.0)  \\
\textit{Specify} & 53.7 ($\uparrow$ 1.4) & 52.2 ($\uparrow$ 2.0) & 45.3 ($\uparrow$ 1.9)  \\
\hline
\end{tabular}
\end{table}

\section{Task 2: Information Retrieval (IR)}
This section aims to investigate the impact of integrating ambiguity into LLM reasoning on IR performance. Table~\ref{tab:ir_results} shows IR results of the two different interaction scenarios (\textit{select} \& \textit{respond}) and the baseline without clarification. We detail the result alongside three successive turns. Generally, we observe that AT-CoT $>$ CoT $>$ AT-standard $\approx$ standard for most of the interaction modes and turns. For instance, clarifications obtained with the method AT-COT allow to reach the best IR metrics values for the TREC Web Track 2013-2014 dataset over all turns (0.397, 0.391, and 0.381 for each turn respectively vs 0.392, 0.384, and 0.373 at most for the baselines). We also note that IR performance is always better for the \textit{respond} interaction mode corresponding to the generation of clarifying questions (in contrast to the \textit{select} mode based on query reformulation. Altogether, these results highlight two main conclusions: 1) it aligns with our remarks on the CG performance, demonstrating the benefits of introducing ambiguity-oriented LLM reasoning for clarification, both intrinsically and extrinsically. And 2) this reinforces our hypothesis based on the need for clarification interactions based on ambiguity and reasoning in IR. Our findings also demonstrate the robustness of our methodology in interaction scenarios. For both interaction scenarios \textit{select} and \textit{respond}, AT-CoT consistently provides the best IR performance, implying that our method can adapt to various real-world scenarios such as query suggestion-based scenarios (e.g. search suggestions) or chat scenarios (e.g. chatbot). 

\begin{table}[t]
\setlength{\tabcolsep}{3.5pt}
\caption{Results on IR datasets based on user simulation. Scores are in nDCG@10 (\%) for Trec Web Track 2009-2012 and TREC Web Track 2013-2014; MRR@10 (\%) for TREC DL Hard. $^{*}$, $^{\dag}$, $^{\Delta}$, indicates statistically significant improvements over standard, AT-standard, CoT respectively with $p < 0.01$ under a t-test.}\label{tab:ir_results}
\centering
\resizebox{\linewidth}{!}{\begin{tabular}{l|c@{}c@{}|c@{}c@{}|c@{}c@{}}
\hline
& \multicolumn{2}{c|}{\textit{TREC Web Track 09-12}} & \multicolumn{2}{c|}{\textit{TREC Web Track 13-14}} & \multicolumn{2}{c}{\textit{TREC DL Hard}} \\
\cline{2-7}
& \textit{select~} & \textit{respond} & \textit{select~} & \textit{respond} & \textit{select~} & \textit{respond} \\
\hline
\textit{w/o clarification} & 0.123 & 0.123 & 0.277 & 0.277 & 0.084 & 0.084 \\
\hline
\multicolumn{7}{c}{\textit{Turn-1}} \\
\hline
standard & 0.161 & 0.232 & 0.336 & 0.387 & 0.060 & 0.120 \\
AT-standard & 0.165 & 0.230 & 0.337 & 0.383 & 0.066 & 0.113 \\
CoT & 0.174$^{*\dag}$ & 0.238 & 0.341 & 0.392$^\dag$ & 0.063 & 0.123$^\dag$  \\
AT-CoT & \textbf{0.188}$^{*{\dag\Delta}}$ & \textbf{0.244}$^{*\dag}$ & \textbf{0.347}$^{*\dag}$ & \textbf{0.397}$^{\dag}$ & \textbf{0.074}$^{*\Delta}$ & \textbf{0.125}$^{\dag}$ \\
\hline
\multicolumn{7}{c}{\textit{Turn-2}} \\
\hline
standard & 0.152 & 0.223 & 0.307 & 0.379 & 0.054 & 0.127 \\
AT-standard & 0.149 & 0.228 & 0.291 & 0.376 & 0.052 & 0.151$^*$ \\
CoT & 0.160$^{*\dag}$ & 0.226 & 0.310$^\dag$ & 0.384 & 0.062$^\dag$ & 0.174$^{*\dag}$  \\
AT-CoT & \textbf{0.176}$^{*\dag\Delta}$ & \textbf{0.233}  & \textbf{0.320}$^{*\dag\Delta}$ & \textbf{0.391}$^{*\dag}$ & \textbf{0.071}$^{*\dag\Delta}~$ & \textbf{0.184}$^{*\dag\Delta}$  \\
\hline
\multicolumn{7}{c}{\textit{Turn-3}} \\
\hline
standard & 0.141 & 0.212  & 0.295 & 0.371 & \textbf{0.056} & 0.141  \\
AT-standard & 0.149 & 0.213  & 0.276 & 0.367  & 0.051 & 0.154  \\
CoT & 0.148 & \textbf{0.216} & 0.300$^\dag$ & 0.373  & 0.054 & 0.184$^{*\dag}$  \\
AT-CoT & \textbf{0.152} & 0.213 & \textbf{0.305}$^\dag$ & \textbf{0.381} & 0.052 & \textbf{0.188}$^{*\dag}$  \\
\hline
\end{tabular}}
\end{table}

\paragraph{Per-turn IR performance.} We observe the same pattern of performance changing across multiple conversation turns for all prompting schemes. For example, under \textit{select}, the IR performance reaches the highest value in the first turn, then monotonically decreases; for Trec DL Hard under \textit{respond}, the IR performance steadily increases as the conversation continues. This IR performance changing pattern is coherent to query difficulties. As a collection that contains complex queries from Trec DL 2019/2020 datasets~\citep{Craswell2019TrecDl,Craswell2020TrecDl}, queries in Trec DL Hard are relatively longer, more challenging in terms of resolving ambiguities. Therefore, Trec DL Hard may necessitate multi-turn conversations to fully clarify ambiguities, which is reflected in the increasing scores across conversation turns. Similarly, for Trec Web Track datasets, the peak IR performance appearing at the first turn is reasonable, since queries in these datasets are not highly ambiguous. Nevertheless, in terms of turn-specific IR performances, AT-CoT still outperforms other prompting schemes, demonstrating that there is no need to increase conversation turns to reflect the improvements of AT-CoT. Regardless of how many turns of conversation a user intends to have, AT-CoT is able to provide better clarifications compared to other prompting schemes.

\section{Task 3: Alignment between Clarification Generation \& Information Retrieval}\label{sec:align_cg_and_ir_tasks}
To mitigate potential bias introduced in user simulation, we further align the performance of CG and IR by using Qulac-Trec Web Track 2009-2012 as mentioned in Section~\ref{sec:cg_ir_datasets}. Since reference CQs in Qulac are only provided for initial queries, we use the CG and IR results from the first turn under \textit{respond}. We compute the Pearson correlation coefficient to measure the strength of the linear relationship between the CG and IR results, and obtain $r=0.92$, $p=0.08$, which shows a strong positive correlation. We hypothesize that the insignificance of this correlation may due to the complexity of the document collection, which is insufficient to differentiate the quality of clarifications. A query may be refined by high-quality CQs through user simulation, but there lack of relevant documents to account for this refinement and reflect it by IR performance. However, given that we obtain a correlation coefficient greater than 0.9, it does not undermine our observation that IR performance is correlated to CG, i.e. IR performance improvements brought by AT-CoT are due to better clarifications. 

\section{Conclusion}
In this work, we investigate the integration of ambiguities and reasoning into LLM prompting methods for clarification, proposing a new action-based ambiguity type taxonomy and a new prompting scheme, AT-CoT. Experiments on clarification generation and information retrieval datasets demonstrate the effectiveness of our methodology. Besides, in-depth analyses show that our method is robust in different clarification interaction scenarios and can capture the clarification needs of datasets with different characteristics. 

However, our work is not without limitations. First, we establish an ambiguity type taxonomy containing three general ATs for integration with LLM reasoning. We do not experimentally study the impact of AT granularity, particularly investigating whether reasoning over a structured ambiguity taxonomy would be beneficial. Second, we only use Llama-3-8B without testing LLMs of different scales. It would be interesting to study the reasoning capability of different model scales. Nevertheless, we believe that our work acts as a foundation to better understand the role of ambiguity types in LLM prompting methods for clarification and may provide useful insights for future work.

\begin{acks}
This work benefited from support from the French National Research Agency (Project GUIDANCE, ANR-23-IAS1-0003) and SCAI (Sorbonne Center for Artificial Intelligence).
\end{acks}


\appendix
\end{sloppypar}
\end{document}